\documentclass[preprint, nobibnotes, aps, superscriptaddress]{revtex4-1}
\usepackage{amsmath}
\usepackage{amssymb}
\usepackage{hyperref}
\usepackage{physics}
\usepackage{graphicx}
\usepackage{color}
\usepackage{subcaption}
\usepackage{siunitx}

\begin{document}

\title{Inverse Design in the Complex Plane: Manipulating Quasi--Normal Modes}

\author{J. R. Capers}
\email{jrc232@exeter.ac.uk}
\affiliation{Department of Physics and Astronomy, University of Exeter, Stocker Road, Exeter, EX4 4QL}

\author{D. A. Patient}
\email{dp348@exeter.ac.uk}
\affiliation{Department of Physics and Astronomy, University of Exeter, Stocker Road, Exeter, EX4 4QL}

\author{S. A. R. Horsley}
\affiliation{Department of Physics and Astronomy, University of Exeter, Stocker Road, Exeter, EX4 4QL}

\date{\today}

\begin{abstract}
    Utilising the fact that the frequency response of a material can be decomposed into the quasi--normal modes supported by the system, we present two methods to directly manipulate the complex frequencies of quasi--normal modes in the complex plane.
    We first consider an `eigen--permittivity' approach that allows one to find how to shift the permittivity of the structure everywhere in order to place a single quasi--normal mode at a desired complex frequency.
    Secondly, we then use perturbation theory for quasi--normal modes to iteratively change the structure until a given selection of quasi--normal modes occur at desired complex frequencies.
\end{abstract} 

\maketitle

\section{Introduction}
    
\begin{figure}
        \centering
        \includegraphics[width=\linewidth]{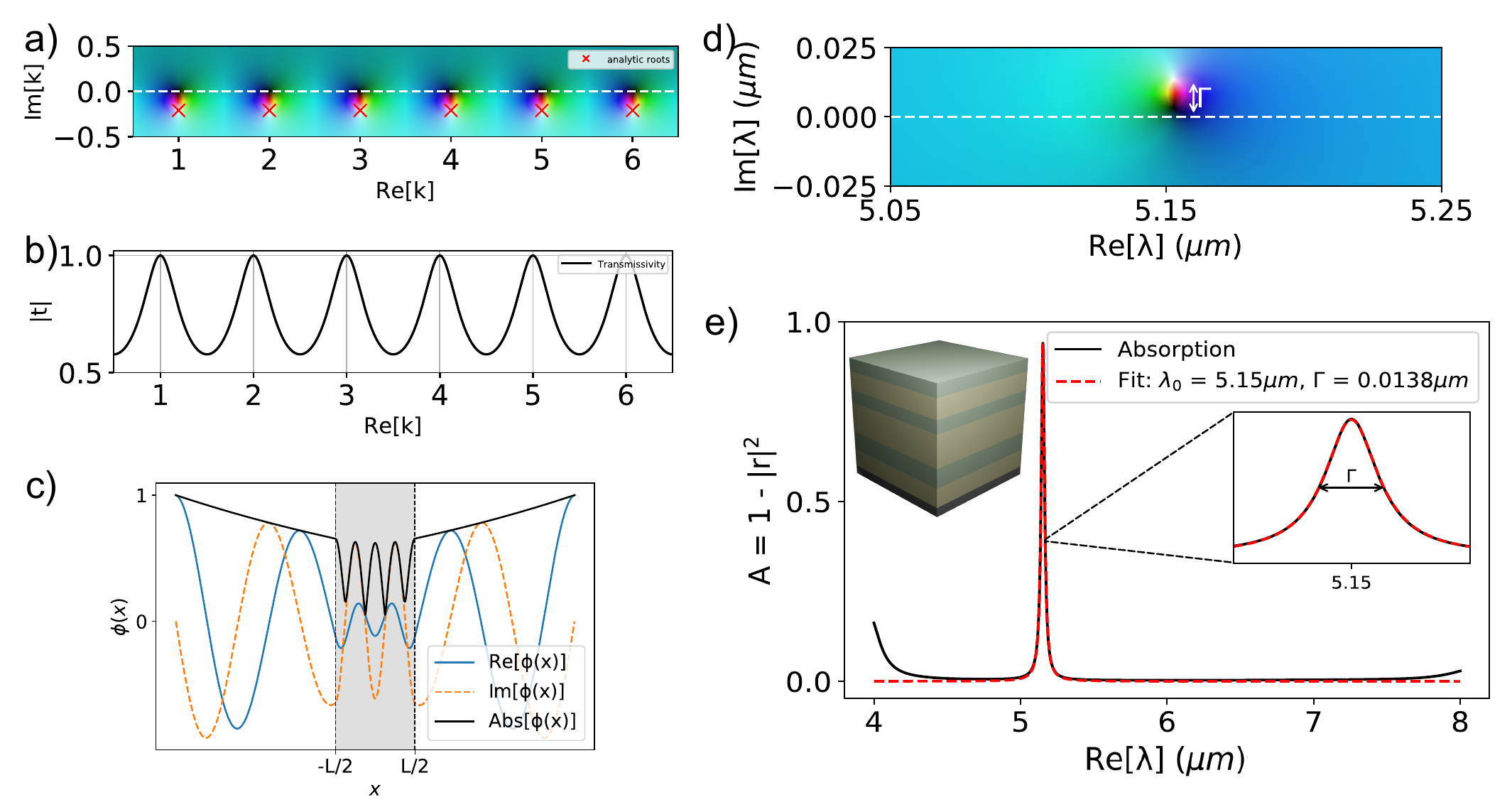}
        \caption{The quasi--normal modes of a dielectric slab (a-c) and the absorbing stack (d-e). 
        The reflection coefficient in the complex plane (a) and transmission ($\sqrt{1 - |r|^2}$) along the real frequency axis (b). 
        The red crosses represent the analytic solution to Eq.~(\ref{eq:AnalyticQNMSlab}) for $n_b=1$, $n_r = \pi$, $L = 1$. 
        The real component of the QNMs are associated with the peaks in transmission. 
        The real (blue), imaginary (orange dashed) and absolute (black) field distribution (c) of the $m = 3$ mode is shown to have the characteristic exponential growth in space. 
        For the (near) perfect absorber (depicted in inset, green layers are Germanium, yellow are Silicon Oxide, with a Tungsten substrate), the complex reflection coefficient (d) shows a single QNM. 
        The absorption spectrum (e) shows a large resonance in the mid-IR, with a resonant frequency ($\lambda_0$) and linewidth ($\Gamma$) directly associated with the QNM, which can be understood in terms of the poles of the associated Lorentzian (red dashed line).}
        \label{fig:AllBase}
    \end{figure}

    Quasi--normal modes (QNMs) are the complex frequency bound states of a system.  
    They were first used in quantum mechanics to describe alpha decay \cite{Gamow1928, Bethe1937}, and have since found utility in modelling radiation in many different systems, from black holes \cite{Chandrasekhar1975}, photonic resonators \cite{Kristensen2020} to leaky waveguides \cite{Ghatak1985, Hu2009}.
    QNMs correspond to the poles of the scattering matrix in the complex frequency plane \cite{Alpeggiani2017, Tikhodeev2017}, where the waves at the boundary of the system are purely out--going.
    The effect of a structured environment can, for example, be analysed by decomposing the Purcell factor in terms of these QNMs \cite{Zschiedrich2018}, and through calculating how small changes in the system perturb the QNMs, deeper insight into sensing has been developed \cite{Yang2015, Both2022}.
    Here, motivated by the connection between the location of poles in the complex plane and physical properties such as transmission, we combine ideas from inverse design with the QNM approach to modelling resonator systems to design materials that have poles at specific complex frequencies.
    Perhaps the simplest example of a system supporting QNMs is a homogeneous dielectric slab (refractive index $n_R$ in some background index $n_B$). 
    For this simple case, the complex frequencies of the QNMs can be found analytically \cite{Chandrasekhar1975, Kristensen2020} as 
    \begin{equation}
        \label{eq:AnalyticQNMSlab}
        k_m L = \frac{2 \pi m + i \ln \left[ \left( n_R - n_B \right)^2/\left( n_R + n_B \right)^2 \right]}{2 n_R},
    \end{equation}
    where $m$ is an integer and $L$ is the width of the slab. 
    Figs.~\ref{fig:AllBase}(a-c) demonstrate that poles in the reflection coefficient as a function of complex $k$ correspond to QNMs, which are in turn associated with peaks in transmission. 
    Examining the field, shown in Fig.~\ref{fig:AllBase}(c), at a complex $k$ value associated with a QNM shows the characteristic exponential growth in space.
    
    More complicated systems can also be understood in terms of QNMs.   For example, multilayer dielectric absorbers (e.g. the mid--infrared absorber presented in~\cite{Sakurai2019}) can be understood this way.  
    The absorption of the structure given in~\cite{Sakurai2019}, along with the reflection coefficient in the complex wavelength plane is shown in Figs.~\ref{fig:AllBase}(d-e).  Fitting the Lorentzian
    \begin{equation}
        \label{eq:Lorentzian}
        \mathcal{L}(\lambda) = \frac{\Gamma}{(\lambda - \lambda_0)^2 + \Gamma^2}
    \end{equation}
    to the absorption peak, we find the peak wavelength is $\lambda_0 = 5.15 \mu$m and the linewidth $\Gamma = 0.0138 \mu$m. 
    This corresponds to a pole of the reflection coefficient in the complex plane  at $\lambda_0 + i \Gamma$, as shown in Fig.~\ref{fig:AllBase}(e).
    
    While QNMs provide a valuable framework to understand resonators, the ability to \emph{design} the spectral response of materials is key to e.g. more efficient photovoltaic cells \cite{Zoysa2012} and sensors \cite{Liu2010}.
    For sensing applications, narrow resonances at particular wavelengths are desirable \cite{Landy2008, Luo2016, Lochbaum2017}, while energy harvesting requires large absorption over a broad band \cite{Aydin2011, Pala2009, Zhou2021, Ding2016}.

    When designing spectral features, one can employ the physical insight provided by QNMs to greatly simplify the problem.
    For example, one way to approach the inverse design problem for absorbers is to try to move the QNM to a desired complex frequency \cite{Grigoriev2013}.
    In this way, one can tailor scattering effects \cite{Wu2020}, design absorbers \cite{Ming2019} and manipulate exceptional points \cite{Yan2020} with minimal numerical complexity.
    To date, however, these approaches address the forwards problem, finding how the pole moves if the resonator geometry is changed.
    We instead solve the inverse design problem of designing materials with poles at specific complex frequencies, using only simple techniques.
    % Throughout this work, results are obtained by either back--integrating the Helmholtz equation using Scipy's `odeint' \cite{Virtanen2020}, or using the transfer matrix \cite{Born2013}, although comparisons with full--wave solutions in COMSOL Multiphysics \cite{COMSOL} are provided.
    
    We present two methods for placing QNM poles at arbitrary complex frequencies.
    Firstly, we re--formulate the eigenvalue problem of the Helmholtz equation to find a complex constant value by which the permittivity of a structure should be shifted to place a pole in the desired location.
    Secondly, we employ QNM perturbation theory to find how to change the spatial distribution of material to move around several poles in the complex frequency plane.
    These methods enable the simultaneous control of resonance wavelength \emph{and} linewidth, for the design of absorbers and sensors. 

\section{Eigen--Permittivities}
    
    One way to \emph{find} the locations of quasi--normal modes (QNMs) is to formulate the Helmholtz equation for the out--of--plane electric field $\phi$, as an eigenvalue problem for complex wave--numbers $k$
    \begin{equation}
        -\frac{1}{\varepsilon (x)} \dv[2]{\phi}{x} = k^2 \phi \label{eq:eigen_eqn_k}.
    \end{equation}
    However, to find the QNMs the correct boundary condition must be imposed on $\phi$.
    Originally derived by Sommerfeld \cite{Sommerfeld_pde}, but since used to model black hole radiation \cite{Zerilli1970, Kapur1938}, the appropriate boundary condition is that the wave is purely outgoing.  
    For example, on the either side of a planar medium,
    \begin{equation}
        \dv{\phi (x)}{x} = \pm ik \phi (x) ,
        \label{eq:outgoing_bc}
    \end{equation}
    as $x \rightarrow \pm\infty$.
    To numerically find the QNMs of our system, we imposed this boundary condition within a finite difference approximation, adapting the elements of the Laplacian at the boundaries e.g. for $N$ points the value of the field at the final point on the right of the system is fixed to be $\phi_{N+1}=\phi_{N}+i k\Delta x \phi_{N}$, giving
    \begin{equation}
    \label{eq:QNMLapMat}
        \dv[2]{\phi}{x} \approx \frac{1}{(\Delta x)^2}
        \begin{pmatrix}
            (ik\Delta x - 1) & 1 & 0 & 0 \\
            1 & -2 & 1 & 0 \\
            0 & 1 & -2 & 1 \\
            0 & 0 & 1 & (ik\Delta x - 1)
        \end{pmatrix}
        \begin{pmatrix}
            \phi_1 \\
            \phi_2 \\
            \phi_3 \\
            \phi_4
        \end{pmatrix} .
    \end{equation}
    It is now evident that solving the eigenvalue problem required to find the QNMs is challenging \cite{Lalanne2019} as the eigenvalue $k^2$ also appears in the boundary condition.
    To avoid solving this non--linear problem, it has recently been noted by Chen et al. \cite{Chen2019} that the analysis of QNMs can be simplified by working in terms of real wave--numbers but extending the \emph{permittivity} into the complex plane.
    Despite the utility of the normal mode framework of Chen et. al. \cite{Chen2019} for employing modal expansions we are trying to engineer the resonance location (related to ${\rm Re}[k]$) and linewidth (given by ${\rm Im}[k]$).
    The location of the QNM frequency trivially encodes these features we are trying to engineer.
    Employing the insight of Chen et. al., we write the permittivity as a spatial variation plus a constant background $\varepsilon (x) = \varepsilon_s (x) + \varepsilon_b$ allows us to recast the Helmholtz equation as an eigenvalue problem for the permittivity
    \begin{equation}
        \label{eq:EigenProblem}
        -\frac{1}{k^2} \left( \frac{d^2}{dx^2} + k^2 \varepsilon_s(x) \right) \phi (x) = \varepsilon_b \phi (x).
    \end{equation}
    Rather than using this to find the QNMs of a system, we show that this can be used to design the complex frequencies of the QNMs.
    
    To do this, we take a known spatially varying permittivity, such as the dielectric step or absorber stack e.g. from \cite{Sakurai2019}, and choose a $k \in \mathbb{C}$ at which we would like a QNM to occur.
    We then numerically solve the eigenvalue problem Eq. (\ref{eq:EigenProblem}) using the finite difference method Eq.~(\ref{eq:QNMLapMat}), along with  standard matrix libraries, to find a complex eigen--permittivity that allows us to form a structure with $\varepsilon (x) = \varepsilon_s (x) + \varepsilon_b$ with a pole at the chosen complex frequency.
    
\begin{figure}
        \centering
        \includegraphics[width = \columnwidth]{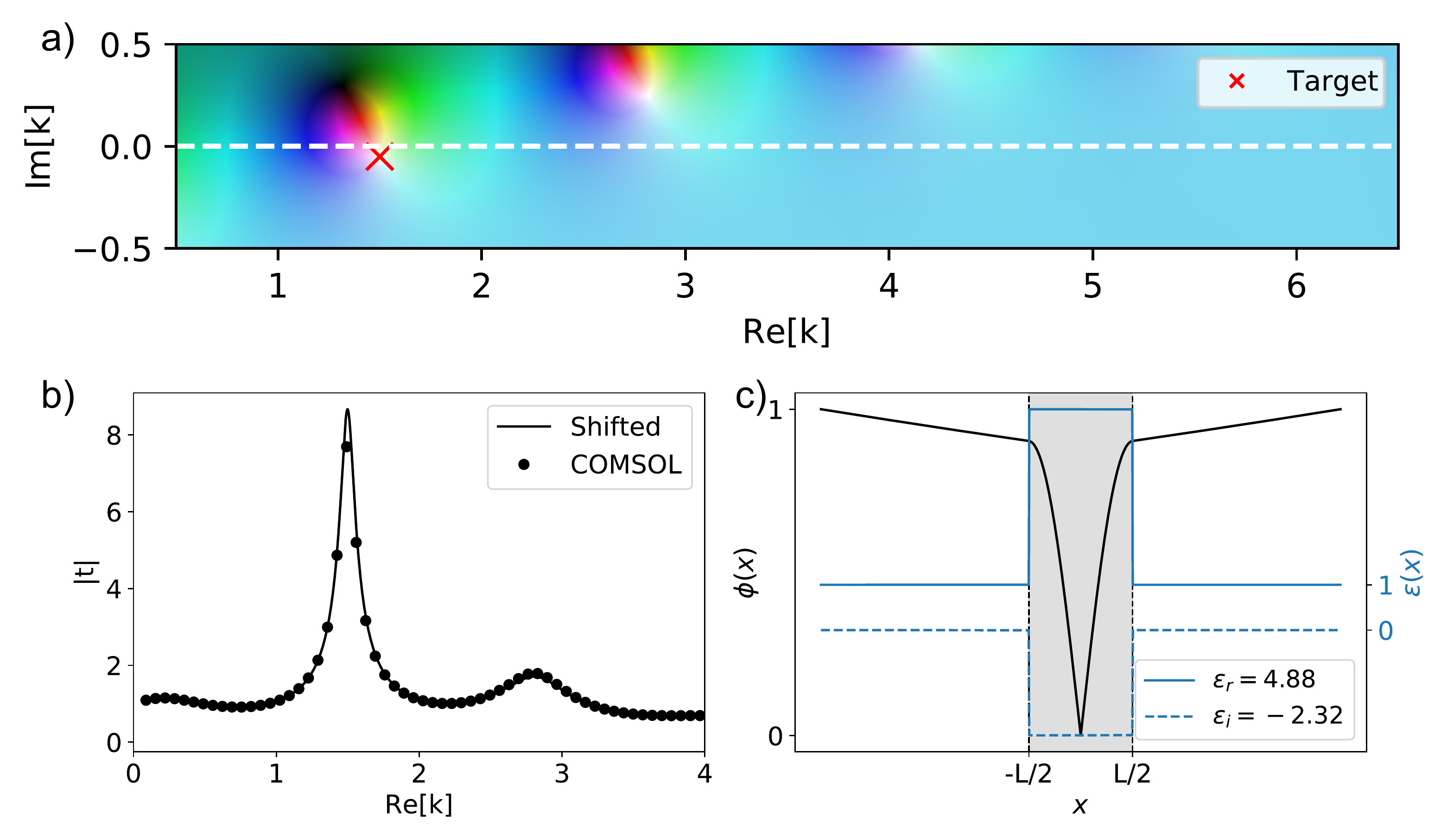}
        \caption{A background permittivity $\varepsilon_b = -4.99-2.32i$ is found as a solution to Eq.~(\ref{eq:QNMLapMat}) which, when combined with the original structure $\varepsilon_s(|x| < L/2) = \pi^2$ will contain a pole at the desired complex frequency of $k = 1.5 - 0.05i$. The reflection coefficient of the new structure is plotted in the complex plane (a). The transmission along the dashed white line, where $\rm{Im}[k] = 0$ is plotted (b), alongside the field distribution plotted at the complex frequency $k$ (c). Overlaid on the transmission calculations are results found using COMSOL Multiphysics \cite{COMSOL}.}
        \label{fig:EigenFields}
    \end{figure}

    We first apply the method to the homogeneous slab. 
    In Fig.~\ref{fig:EigenFields} we design the new structure to support a QNM at the frequency $k = 1.5 - 0.05 i$. 
    For the $N \times N$ Laplacian matrix, there are $N$ possible values for $\varepsilon_b$ that will satisfy this condition. 
    Taking the lowest absolute valued background (to minimise numerical error) permittivity $\varepsilon_b = -4.99 - 2.32 i$, we find that the new structure now supports a QNM at our chosen $k$.
    This is shown in Fig.~\ref{fig:EigenFields}(a). 
    The transmission, Fig.~\ref{fig:EigenFields}(b), shows a large peak at the real frequency associated with the QNM and has values $|t|>1$ due to the gain that has been added to the system. 
    Although the location of the pole can be manipulated solely by changing the height of the barrier, in order to manipulate the real and imaginary parts independently, control over both the real and imaginary permittivity is required. 
    As might be anticipated, in order to move a pole closer to the real frequency axis, without changing the resonant frequency, gain is required.  
    Conversely, loss is required to move the pole further away from the real axis.
    The field profile, shown in Fig.~\ref{fig:EigenFields}(c), still has the exponential growth characteristic of QNMs.
    
\begin{figure}
        \centering
        \includegraphics[width = 0.75\columnwidth]{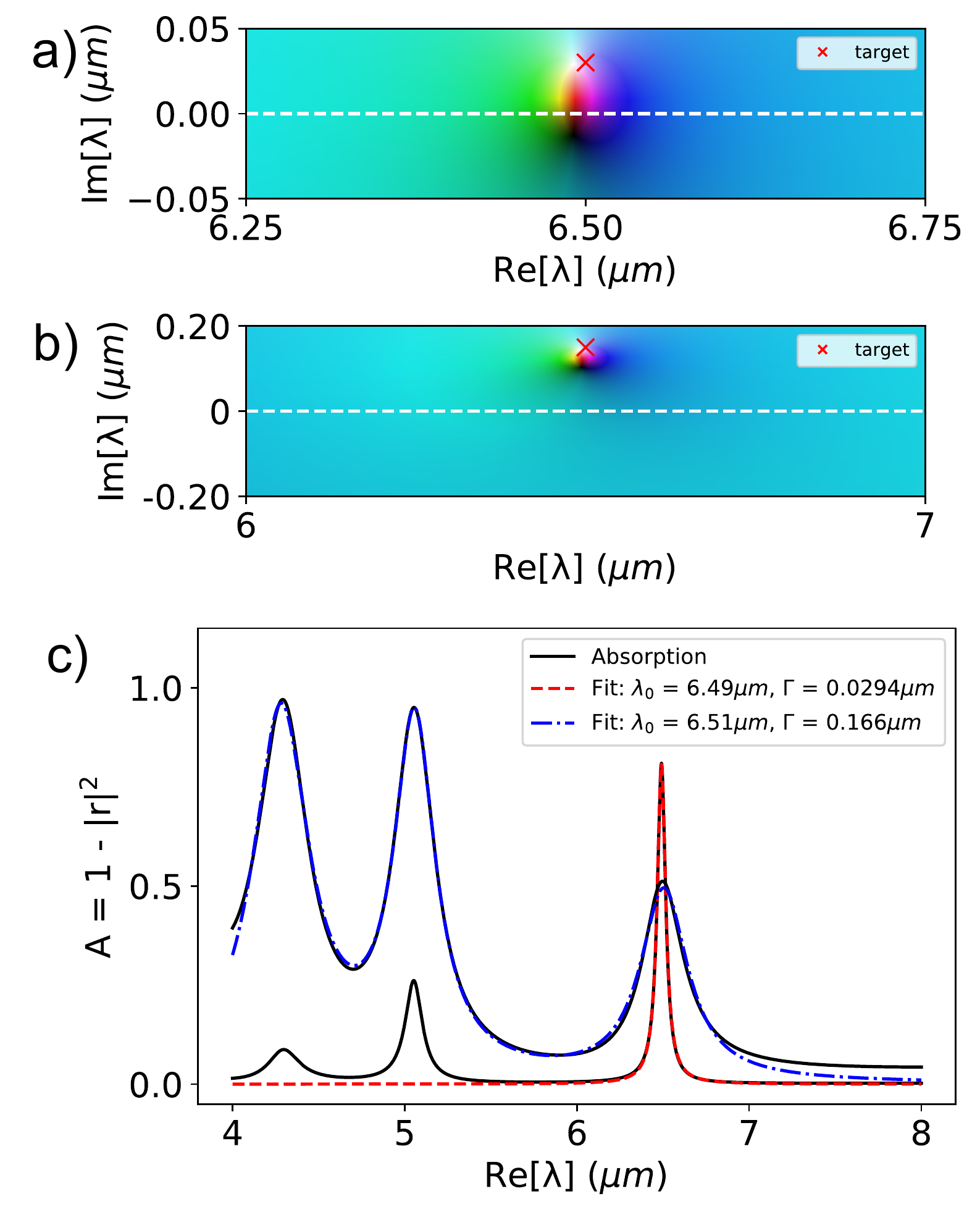}
        \caption{The original absorbing stack, shown in Fig. \ref{fig:AllBase}(e) has been modified into two structures that contain a QNM at $\lambda_1 = (6.5 + 0.03 i)\mu {\rm m}$ and $\lambda_2 = (6.5 + 0.15i) \mu {\rm m}$ respectively. The former is close to the real axis, corresponding to a narrow bandwidth, while the latter has a broader bandwidth. Plotted on (a) and (b) respectively are the reflection coefficients in the complex plane, showing that a QNM is indeed located at the chosen complex frequency. The absorption spectra of the two structures are plotted as a function of real wavelength (c). Fitted Lorentzians in dashed red (blue) correspond to fitting to the narrow (broad) resonance, verifying the complex frequencies of the QNMs. For the broadband case, we must fit a sum of 3 Lorentzians to accurately model the spectral profile, and obtain the correct fitting parameters.}
        \label{fig:SakuraiShifted}
    \end{figure}

    Next, we apply the same eigen--permittivity method to the absorbing stack shown in Fig.~\ref{fig:AllBase}(e).
    For this structure, we must take care that the correct boundary conditions are imposed.
    The opaque metal substrate requires the Dirichlet boundary condition $\phi = 0$, while the outgoing wave boundary condition must be imposed at the top of the stack.
    Choosing two target bandwidths, for the same resonance wavelength, $\lambda_1 = (6.5+0.03i)\mu$m and $\lambda_2 = (6.5+0.15i)\mu$m, we obtain background permittivities of $\varepsilon_{b,1} = 3.27 - 0.01i$ and $\varepsilon_{b,2} = 3.28 + 0.29i$.
    The effect of the background shift on the pole locations is shown in Fig.~\ref{fig:SakuraiShifted}(a-b). 
    Accordingly, the poles are found at the expected complex frequencies. The absorption, shown in Fig.~\ref{fig:SakuraiShifted}(c) plotted along the white dashed line ($\rm{Im}[\lambda] = 0$) is also provided, with a fitted Lorentzian to extract the properties of the resonances and verify that it corresponds to the QNM frequencies.
    
    We can also apply this design procedure to impose the condition of coherent perfect absorption (CPA) at a given complex frequency.  This can be understood as the time reverse of QNMs \cite{Chong2010} were the wave is purely incoming rather than outgoing.
    The wavelengths at which a structure behaves as a perfect absorber are related to the locations of zeros on the real axis, rather than poles.
    With our eigen-permittivity formulation, we can find the background permittivity value required to make the device a perfect absorber at a frequency of choice.
    To do this, we simply take the outgoing boundary condition Eq. (\ref{eq:outgoing_bc}) and replace it with the incoming boundary condition
    \begin{equation}
        \label{eq:incoming_bc}
        \frac{d \phi(x)}{dx} = \mp i k \phi(x)
    \end{equation}
    as $x \rightarrow \pm \infty$. 
    This changes the boundary elements in the Laplacian Eq.~(\ref{eq:QNMLapMat}) from $i k \Delta x - 1$ to $-i k \Delta x - 1$. 
    
\begin{figure}
        \centering
        \includegraphics[width=\linewidth]{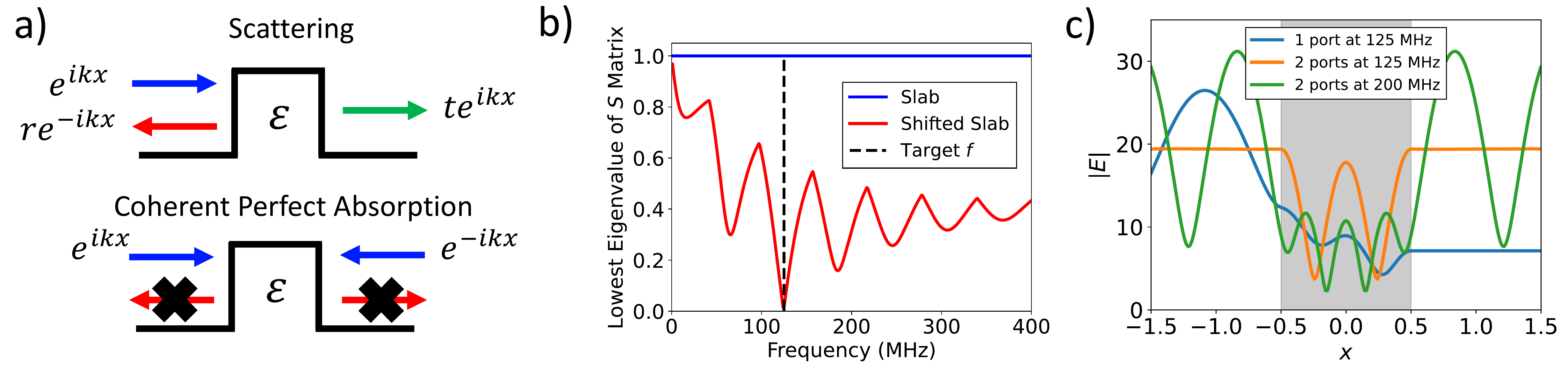}
        \caption{An example of using our eigen--permittivity method to design a structure that exhibits coherent perfect absorption, shown schematically in a).
        Under incidence from one direction, the structure scatters in the usual way, but under incidence from both sides reflection vanishes.
        We design a permittivity step of length $L=1$ of permittivity $\varepsilon = \pi^2 + (-3.98+i \ 1.59)$ that exhibits this behaviour at the desired frequency of 125 MHz.
        To verify this, we show b) the smallest eigenvalues of the scattering matrix of the structure.
        Vanishing eigenvalue indicates coherent perfect absorption.
        This can be clearly observed at the target frequency of 125 MHz.
        The fields c), also indicate coherent perfect absorption.
        Under excitation from one side or off of the target frequency, reflections are observed.
        At the design frequency, there is a standing wave.}
        \label{fig:CPA}
    \end{figure}

    Applying the above changes to the Laplacian matrix, we can take e.g. a slab of dielectric, and rather than choose a complex frequency, pick a real frequency that we wish CPA to occur at. 
    We take a dielectric slab of length $L = 1$ and initial permittivity $\pi^2$ and choose the arbitrary CPA frequency to be 125 MHz.
    The resulting background permittivity required is $\epsilon_b = -9.55+i0.63$. 
    To verify that there is coherent perfect absorption at the chosen frequency, we construct the scattering matrix for the slab under incidence from the left and right side
    \begin{equation}
        \begin{pmatrix}
            \phi_L^{\rm scattered} \\
            \phi_R^{\rm scattered}
        \end{pmatrix}
        = 
        \begin{pmatrix}
            r_L & t_R \\
            t_L & r_R
        \end{pmatrix}
        \begin{pmatrix}
            \phi_L^{\rm in} \\
            \phi_R^{\rm in} 
        \end{pmatrix} ,
    \end{equation}
    noting that CPA occurs when an eigenvalue of the scattering matrix goes to zero \cite{Chong2010}.
    The scattering matrix can be constructed analytically from the transfer matrix or found numerically in full--wave solvers such as COMSOL \cite{COMSOL}.
    In Fig. \ref{fig:CPA} we plot the smallest eigenvalue of the scattering matrix of the slab as a function of frequency.
    A clear dip is seen at the desired frequency.
    We also show field profiles both under incidence from only one side and from both sides at different frequencies.
    Under incidence from only the left side, one can see the usual interference between reflected and incident field to the left of the slab and the constant transmitted field.
    Under excitation from both sides, but away from the target CPA frequency one can see reflection from both sides.
    At the target CPA frequency of 125 MHz, an almost constant field amplitude is observed, indicating perfect absorption.
    
\begin{figure}
        \centering
        \includegraphics[width=\linewidth]{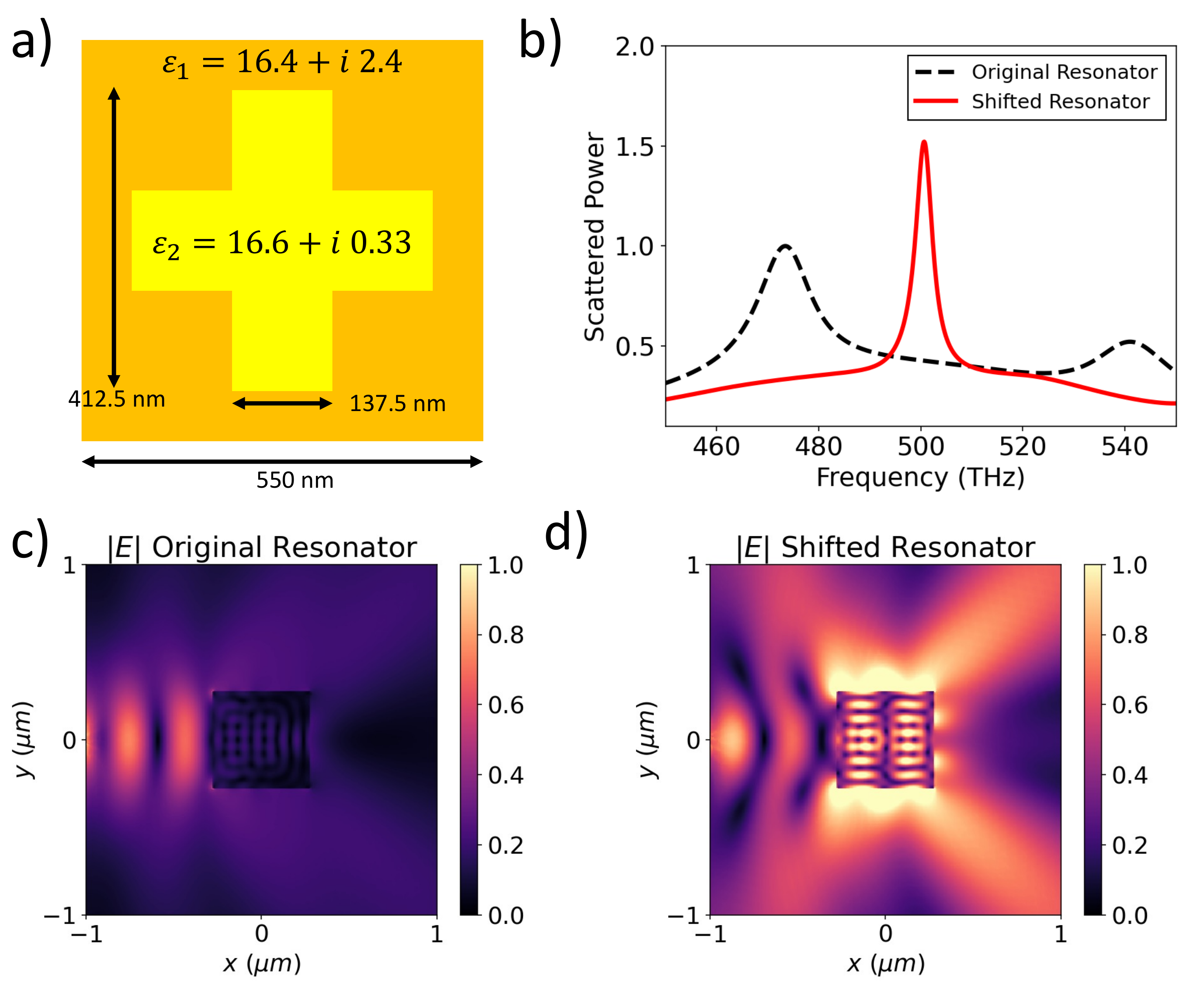}
        \caption{An example of using our eigen--permittivity framework to place the quasi-normal modes of a 2D resonator.
        The resonator, shown inset in a), is made of two different permittivities, $\varepsilon_1$ (silicon at 550 nm) and $\varepsilon_2$ (gallium arsenide at 550 nm).
        We apply our framework to find a permittivity offset to move a pole to the complex frequency (500 + 1i) THz.
        The background is $\varepsilon_b = -1.37 + i 0.88$.
        To verify the location of the pole, we excite the resonator with a point electric dipole, located at (-1100 nm, 0), and calculate the total scattered power, shown in b).
        A clear peak is present in the spectrum of the shifted structure at the desired frequency of 500 THz, which is not present in the spectrum of the un--shifted structure.
        Examining the fields of the resonator driven by a nearby dipole at a frequency of 500 THz, c) and d), we see that the excitation of the mode in the sifted structure greatly increase the scattering.}
        \label{fig:2DQNM}
    \end{figure}

    So far, all of the examples provided have been in 1D. 
    However our method is straightforwardly extended to higher dimensions. 
    To illustrate this we consider a 2D square dielectric resonator, shown in Fig. \ref{fig:2DQNM}(a).
    The resonator is a silicon cross inside a gallium arsenide square.
    To find how to change the permittivity to place a pole at a particular complex frequency, we must solve the eigenvalue problem  Eq. (\ref{eq:EigenProblem}) in 2D.
    To do this, we use COMSOL's coefficient form PDE interface, which allows one to solve problems of the form 
    \begin{equation}
        \lambda^2 e_a \phi - \lambda d_a \phi + \nabla \cdot \left( -c \nabla \phi - \alpha \phi + \gamma \right) + \beta \nabla \phi + a \phi = f ,
    \end{equation}
    where $\lambda$ is the eigenvalue.
    Choosing the coefficients to be $e_a = 0, c = 1, d_a = 1, a = -k^2 \varepsilon$, this becomes exactly the eigenvlaue problem we would like to solve
    \begin{equation}
        \nabla^2 \phi + k^2 \varepsilon \phi = - \lambda \phi ,
    \end{equation}
    where $\lambda = k^2 \varepsilon_b$.
    The outgoing wave boundary condition can be applied to the outside edge of the resonator using the `flux/source' boundary condition.
    Generally, this boundary condition is
    \begin{equation}
        -\boldsymbol{n}\cdot \left( -c \nabla \phi + \alpha \phi + \gamma \right) = g - \phi u,
    \end{equation}
    where $\boldsymbol{n}$ is a unit--vector normal to the surface of the resonator at a given point.
    It is not necessary for $\boldsymbol{n}$ to be normal to the surface, it only needs to point outwards.
    With our parameter choices this becomes 
    \begin{equation}
        \boldsymbol{n}\cdot\nabla \phi = -q \phi .
    \end{equation}
    Setting $q = i k$ gives the correct out--going boundary condition.
    Solving this eigenvalue problem for the 2D geometry shown in Fig. \ref{fig:2DQNM}(a), and choosing the location of the pole to be $f = 500 + 1i$ THz, we find a background permittivity of $\varepsilon_b = -1.37 + i 0.88$.
    To verify that a QNM is now located at the correct complex frequency, we excite the resonator with a nearby point dipole and examine the total scattered power before and after the permittivity shift is applied.
    This is shown in Fig. \ref{fig:2DQNM}(b).
    Once the shift is applied, there is a clear peak in scattered power at the desired wavelength.
    Additionally, the fields when the resonator is driven at 500 THz are shown in Fig. \ref{fig:2DQNM}(c-d).
    Once the permittivty of the resonator is shifted, scattering at the desired frequency is greatly enhanced by the presence of the QNM.
    
    Although simple to implement, this eigen--permittivity method only allows you to choose the complex frequency of a single QNM. We now explore the possibility of applying an iterative method to move one or more QNMs to desired complex frequencies, by changing the spatial variation of the permittivity profile. 
    
\section{Optimisation Approach to Moving Poles}
    
    The second method we present to move quasi--normal modes (QNMs) to desired complex frequencies is to use an iterative procedure, based on perturbation theory.
    Standard Rayleigh--Schrödinger perturbation theory \cite{LL3} of Hermitian quantum mechanics connects a change in the potential $\delta V$ to a change in the $n^{\rm th}$ energy level $E_n$ through the matrix element 
    \begin{equation}
        \delta E_n = \langle \phi_n | \delta V | \phi_n \rangle ,
        \label{eq:RS_perturb}
    \end{equation}
    where the states are normalised so that $\langle \phi_n | \phi_m \rangle = \delta_{nm}$.
    Usually the perturbation to the potential is known and the energy level shifts are calculated (e.g. in the textbook analysis of the Stark effect \cite{LL3} \S 76).
    Being able to analytically connect structure and function is the key to inverse design, allowing one to find derivatives of a quantity of interest (here, the energy) in terms of derivatives of the structure (the potential).
    With this observation, it is possible to use perturbation theory backwards to find how one should change the potential to get a particular energy level.
    This idea can be extended to move the complex frequency of a QNM of an electromagnetic resonator.
    Instead of a potential, we seek to design a permittivity profile $\varepsilon (x)$ that has a QNM, $k_n$, at a particular complex frequency.
    However, as QNMs grow in space, they cannot be normalised. The expressions that connect a change in the permittivity profile to a change in the complex wave--number $k_n$ requires some modification.
    Regularisation techniques have been used to develop a perturbation theory for QNMs in both quantum mechanics \cite{ZelDovich1961, Leung1998} and electromagnetism \cite{Muljarov2010}.
    Perturbation theory can be used to connect a change in the permittivity $\delta \varepsilon (x)$ to a change in the complex frequency of the QNM \cite{Perelomov1998} through
    \begin{equation}
        \delta k_n = \frac{1}{2k_n} \frac{ \int_{-L/2}^{L/2} \phi_n^2 (x) \delta \varepsilon (x) dx }{\langle \phi_n | \phi_n \rangle} ,
        \label{eq:deltak}
    \end{equation}
    where $k = k' + i k''$ and the inner product is now \cite{Leung1998}
    \begin{equation}
        \langle \phi_n | \phi_n \rangle = \int_{-L/2}^{L/2} \phi_n^2 (x) dx + i\left[ \phi_n^2 (-L/2) + \phi_n^2 (L/2) \right] .
    \end{equation}
    If we change the permittivity by a small amount $\Delta \varepsilon$ at a particular location $x_i$ so that $\delta \varepsilon (x) = \Delta \varepsilon \delta (x-x_i)$, we find that 
    \begin{equation}
        \delta k_n = \frac{1}{2k_n} \frac{ \phi_n^2 (x_i) \Delta \varepsilon }{\langle \phi_n | \phi_n \rangle} .
    \end{equation}
    As this is true for all $x_i$, we can divide by the small change in permittivity to find the gradient of the wave--number with respect to the permittivity 
    \begin{equation}
        \pdv{k_n}{\varepsilon} = \frac{\phi_n^2 (x) }{2k_n \langle \phi_n | \phi_n \rangle} . 
        \label{eq:k_grad}
    \end{equation}
    Importantly, this gives a continuous function for the derivative of the complex frequency of the QNM with respect to the spatial structure of the permittivity.
    For example, say we would like to move mode $k_n$ to the complex frequency $k_\star$.
    We can write a suitable figure of merit and it's derivative as
    \begin{align}
        \mathcal{F} &= (k_n - k_\star)^2 , \\
        \pdv{\mathcal{F}}{\varepsilon} &= 2 (k_n - k_\star) \pdv{k_n}{\varepsilon}. \label{eq:update_gradient}
    \end{align}
    Updating the permittivity from iteration $i$ to $i+1$ is done according to 
    \begin{equation}
        \varepsilon^{(i+1)} (x) = \varepsilon^{(i)} (x) + \gamma \pdv{\mathcal{F}}{\varepsilon} ,
        \label{eq:permittivity_update}
    \end{equation}
    where $\gamma$ is the step size.  This makes the evaluation of the figure of merit gradients extremely efficient, similar to the adjoint method \cite{Lalau-Keraly2013}.  Combining this with gradient descent optimisation \cite{Press2007}, we have found how to update the permittivty distribution in order to arbitrarily change the complex frequencies of the QNMs.
    
    \begin{figure}
        \centering
        \includegraphics[width=\linewidth]{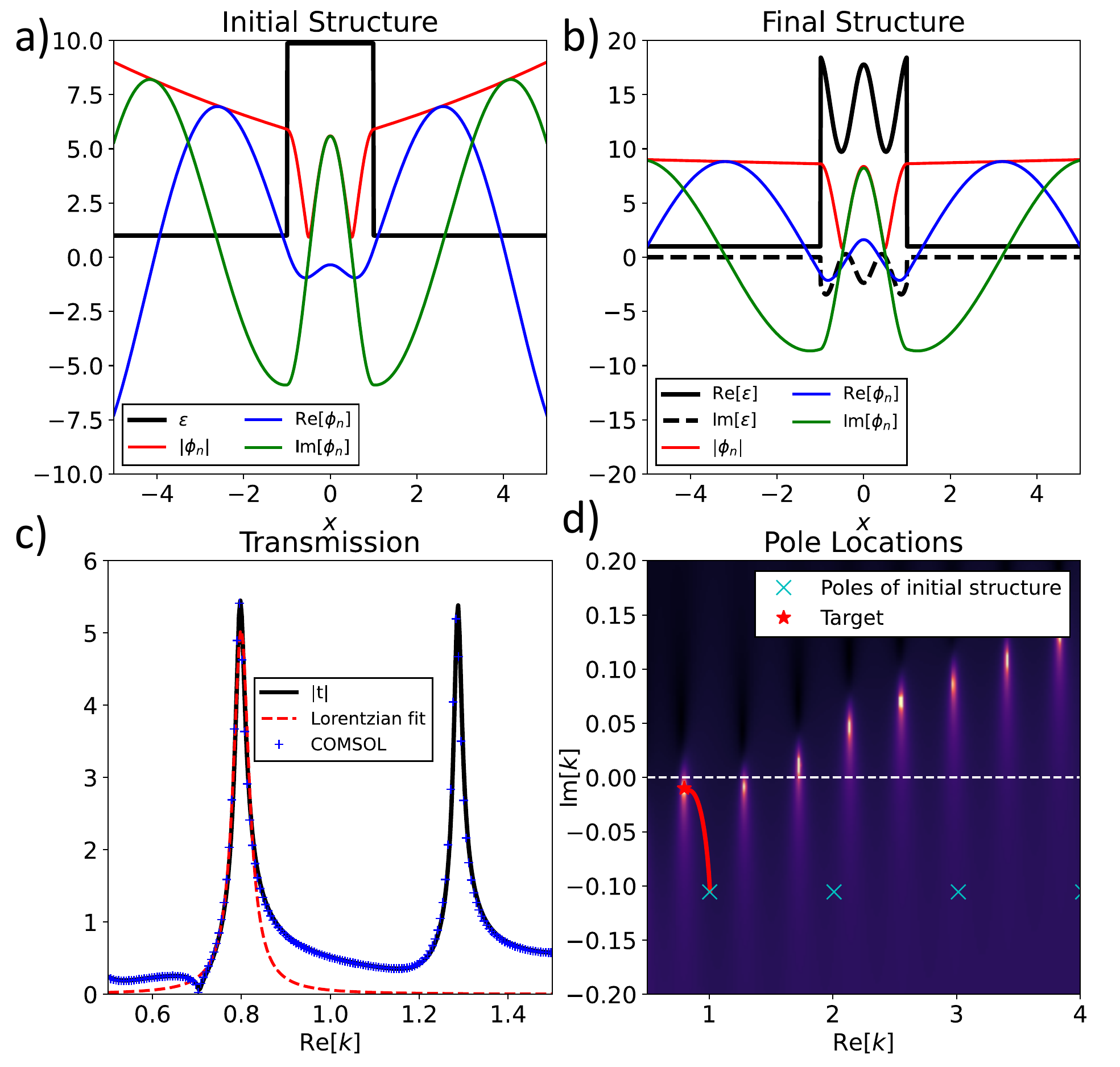}
        \caption{An example of using our iterative method to move a pole to a desired location.
        Beginning from a) a step of dielectric which supports a QNM at $k = 1 - 0.1i$, our iterative method designs the permittivity distribution shown in b), which supports a QNM at the desired frequency $k_\star = 0.8 - 0.01i$.
        The resulting transmission of the structure is shown in c), and compared to a full--wave solver.
        Fitting a Lorentzian to the transmission peak associated with $k_\star$, we extract find that the peak is at $k_0 = 0.799$ with width $\Gamma = 0.0109$.
        The path of the pole over the optimisation is shown in d).}
        \label{fig:moving_pole}
    \end{figure}
    
    An example of this procedure is shown in Fig. \ref{fig:moving_pole}.  We begin by selecting a QNM of the system: the frequency of which we want to modify. The complex wave--number of this mode can be found by root--finding in the complex plane, using i.e. Newton's method.
    Specifying a target frequency of the pole $k_\star$, then using Eqns. (\ref{eq:k_grad}, \ref{eq:update_gradient}, \ref{eq:permittivity_update}) to iteratively update the permittivity distribution allows the pole to be moved to the desired complex frequency.
    At every iteration, $\phi_n$ and $k_n$ must be re--calculated.
    In the example of Fig. \ref{fig:moving_pole} we move the pole originally at $k = 1-0.1i$ to $k_\star = 0.8 - 0.01i$, and show that yields a structure with a peak in transmission at the designed frequency with the designed width.
    It should be noted that while we can move the pole to an arbitrary complex frequency, complete control of both the real and imaginary parts of the permittivity is required.
    
\begin{figure}[h!]
        \centering
        \includegraphics[width=\linewidth]{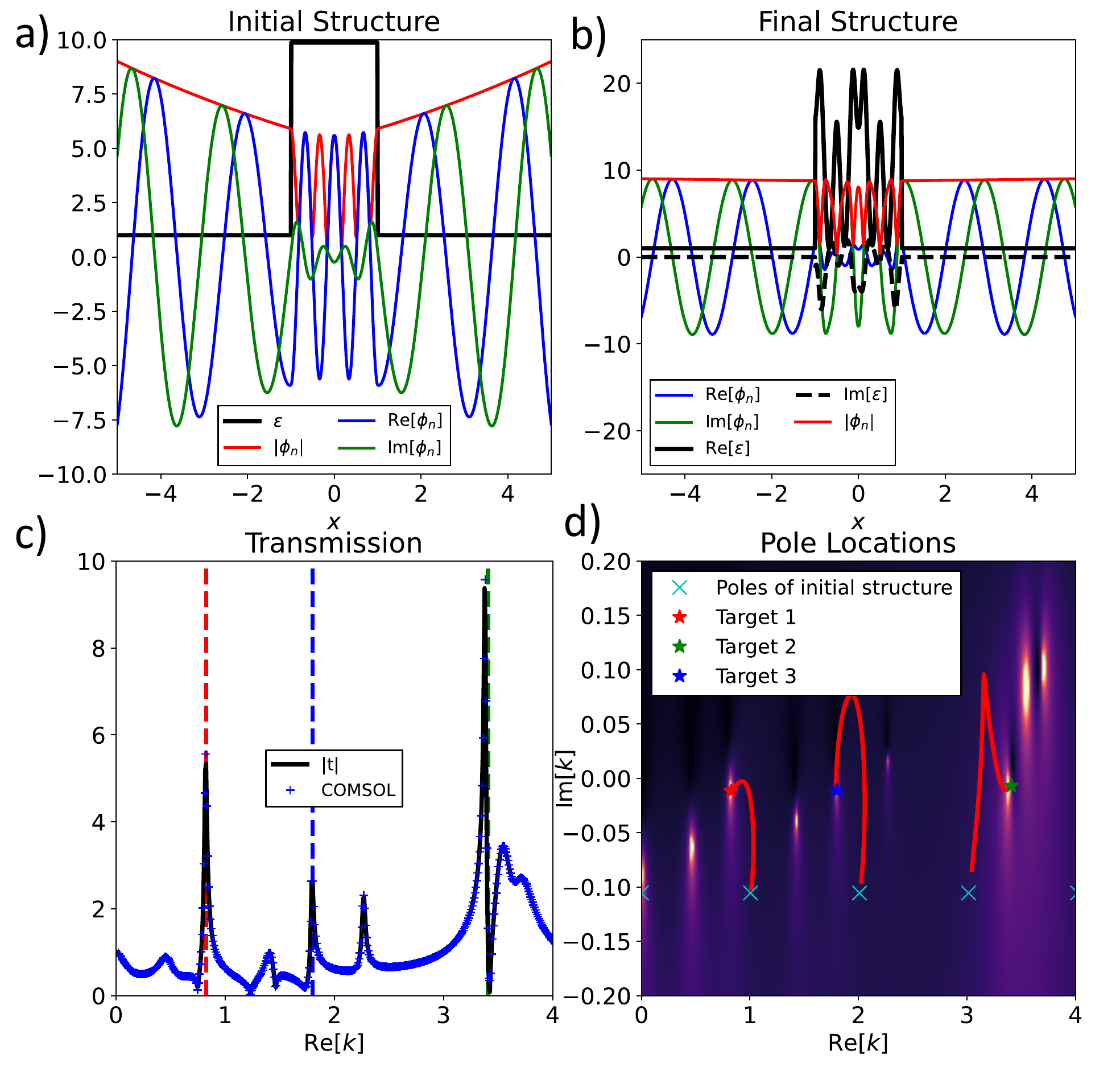}
        \caption{An example of using the iterative method we present to move 3 poles to desired complex frequencies at the same time.
        Beginning from a permittivity step shown in a), the real poles associated with ${\rm Re}[k] = 1,2,3$ are moved to the targets: $k_1 = 0.8-0.007i$, $k_2 = 3.5-0.008i$ and $k_3 = 1.8-0.009i$.
        The resulting permittivity profile is shown in b) and its transmission coefficient in c).  
        Clear peaks are seen at the three target values of $k$.
        The path of the poles over the optimisation is shown in d).}
        \label{fig:moving_3_poles}
    \end{figure}

    As another example of this method, we consider trying to move several poles simultaneously.  
    In Fig. \ref{fig:moving_3_poles} we take the poles originally at $k = 1-0.1i$, $2-0.1i$ and $3-0.1i$ and move them to three different values $k_1, k_2$ and $k_3$.
    Interestingly, due to the presence of other nearby poles, the transmission profile of the resulting structure becomes more complex, however a clear narrow transmission peaks associated with $k_1, k_2$ and $k_3$ are evident.  
    If one controls all poles of interest over a given range of $k$ values, almost complete control over the transmission profile can be obtained.
    
\section{Conclusions and Outlook}

    In this work we address the inverse design problem: `how should one change a photonic system to ensure a quasi--normal mode appears at a pre-determined complex frequency?'. 
    We propose two approaches to answer this question.
    The first is to re-express the permittivity of a system as the original permittivity profile, plus some global background shift $\varepsilon_s(x) + \varepsilon_b$. 
    This allows us to write the Helmholtz equation as an eigenvalue problem for the background permittivity $\varepsilon_b$.
    By choosing a target complex frequency, we can find a (complex) background permittivity that can be added to the structure so that a QNM occurs at the desired complex frequency with the desired linewidth.
    This method could be used to modify existing structures to control the frequency and bandwidth of a resonant system.
    We also show that we can apply this method in order to construct materials that, at a single frequency of operation, act as coherent perfect absorbers.
    
    The second approach we develop is an iterative procedure based on perturbation theory: a small change in permittivity can be connected to the shift in complex frequency of a QNM. 
    By defining a suitable figure of merit, and combining with gradient descent optimisation, we can iteratively change the spatial permittivity profile to move a QNM closer to a target frequency. 
    This procedure can also be used to move multiple QNMs to different target frequencies. 
    This iterative approach can be further modified in many ways.
    For example by restricting the search space of $\delta \varepsilon$ to only allow loss rather than gain, or to ensure $\varepsilon(x) > 1$.
    Also, rather than manipulating the full spatial form of $\varepsilon$, we could seek to change only a few free parameters such as width and height of the dielectric step.
    
    The approaches we have developed open up several avenues of exploration to design, for example, broadband absorbers for solar cells and thermal emitters.
    Since the framework also applies to leaky waveguides, our methods could also be used to design leaky--wave antennas.
    Rather than manually changing structural parameters until a QNM appears at the correct complex frequency, the methods we present leverage the benefits of inverse design to rapidly design materials that have the desired properties.
    Importantly, as our methods allow QNMs to be placed exactly, both resonance frequency and linewidth can be tuned with a high degree of accuracy.

\section*{Acknowledgements}

    The authors would like to thank Josh Glasbey and Ian Hooper for many illuminating discussions and Jake Binsley for his assistance with Blender.
    
    We acknowledge financial support from the Engineering and Physical Sciences Research Council (EPSRC) of the United Kingdom, via the EPSRC Centre for Doctoral Training in Metamaterials (Grant No. EP/L015331/1).  J.R.C also wishes to acknowledge financial support from Defence Science Technology Laboratory (DSTL).  S.A.R.H acknowledges financial support from the Royal Society (URF\textbackslash R\textbackslash 211033).  All data and code created during this research are openly available from the corresponding authors, upon reasonable request.

\bibliographystyle{unsrt}
\bibliography{refs}

\end{document}